 \documentclass[final,1p,times]{elsarticle} 

 \usepackage{graphics}
 \usepackage{graphicx}
\usepackage{epsfig}

\usepackage{amssymb}
\usepackage{amsthm}

\journal{Nuclear Instruments and Methods in Physics Research A }

\begin{document}
\begin{frontmatter}


\ead{barnai@wigner.mta.hu}

\title{Pre-Excitation Studies for Rubidium-Plasma Generation}


\author{M. Aladi$^1$, J.S. Bakos$^1$, I.F. Barna$^{*,1,2}$, A. Czitrovszky$^{1}$, G.P. Djotyan$^1$, P. Dombi$^{1}$, D. Dzsotjan$^1$,  I.B. F\"oldes$^1$, G. Hamar$^1$, P. N. Ign\'acz$^1$, M. A. Kedves$^1$, A. Kerekes$^1$,  P. L\'evai$^1$, I. M\'arton$^1$, A. Nagy$^1$, D. Oszetzky$^1$,  M.A. P\'ocsai$^1$, P. R\'acz$^{1,2}$, B. R\'aczkevi$^1$, J. Szigeti$^1$,   Zs. S\"orlei$^1$, R. Szip\H{o}cs$^1$, D. Varga$^1$, K. Varga-Umbrich$^1$, S. Varr\'o$^1$,  L. V\'amos$^1$ and Gy. Vesztergombi$^1$ }

\address{1) Wigner Research Center of the Hungarian Academy of Sciences \\
1121 Budapest, Konkoly Thege \'ut 29-33, Hungary \\
Tel: +36-1-392-2222/3504, Fax: +36-1-392-2598  \\
2) ELI-HU Nonprofit KFT, H-6720 Szeged, Dugonics t\'er 13., Hungary}

\begin{abstract}

The key element in the Proton-Driven-Plasma-Wake-Field-Accelerator (AWAKE) project is the generation of highly uniform plasma from Rubidium vapor.
The standard way to achieve full ionization is to use high power laser which can assure the over-barrier-ionization (OBI) along the 10 meters long active region.
The Wigner-team in Budapest is investigating an alternative way of uniform plasma generation. The proposed Resonance Enhanced Multi Photon Ionization (REMPI) 
scheme probably can be realized by much less laser power. In the following the resonant pre-excitations of the Rb atoms are investigated, 
theoretically and the status report about the preparatory work on the experiment are presented.

\end{abstract}


\end{frontmatter}



\section{Introduction}

After building LEP and LHC, and planning ILC the size and cost of 
classical high energy particle accelerators are reaching their limits 
around 50 MeV/m.

In their historical article \cite{tos} in 1979 T. Tajima and J. M. Dawson
predicted: "glass lasers of power density $10^{18} W/cm^2$ shone on plasmas of 
densities $10^{18} cm^{-3}$ can yield gigaelectronvolts of electron energy per 
centimeter of acceleration distance." With the availability of high power 
short pulse lasers their ideas are becoming reality.

It turned out that it is possible to build not only Laser Wake Field Accelarator (LWFA)  but 
also Particel Driven Wake Field Accelaratior (PWFA). In the laboratories 100 GV/m 
accelerating gradient was demonstrated with laser driven systems \cite{leh} and 
50 GV/m with electron driven case in SLAC  \cite{SLAC}.

At CERN the AWAKE collaboration \cite{awake} has been formed in order to 
demonstrate proton driven plasma wake field acceleration for the first time, 
where the Wigner team is interested in the creation of high uniformity 
plasma which is required to reach the plasma frequency stability at the 
percent level. 

The SPS proton beam in the CNGS facility will be injected into a 10 m 
plasma cell where the long proton bunches will be modulated into 
significantly shorter micro-bunches. These micro-bunches will then 
initiate a strong wakefield in the plasma with peak fields above 1 GV/m.
Though this peak field is much less than the one achieved by laser or 
electron driven systems, the future accelerator based on this technology 
can have a much reduced length compared to proposed linear accelerators.

In the proposal  \cite{prop} a uniform plasma was assumed to be created by 
high enough laser power with over-barrier-ionization (OBI), achieving saturation 
along the whole length of the 10 meter plasma.  In this paper we propose a much softer method by 
applying the so-called  Resonance Enhanced Multi Photon Ionization (REMPI) 
scheme. It is a three photon process which requires significantly lower laser power than the non-resonant OBI process. 
From this three photon ionization process, we investigate the two-photon resonant 
excitation process  with theoretical and experimental method.  


\section{The experimental setup} 

In order to produce a desirable homogeneous laser plasma for the wake field acceleration of particles the ionization process by high intensity laser pulses has to be studied. Therefore the aim of the first experiment is the investigation of high field ionization gain from rubidium atomic beam by using the commercial regenerative Ti:sapphire based chirped-pulse amplifier laser system (Coherent Legend Elite HE-USP). The amplifier delivers 35 fs pulses at 800 nm central wavelength with 1 kHz repetition rate and with 4 mJ pulse energy. The characteristic noise ratio of the average power is smaller than 0.5 \%. The maximum achievable focused peak intensity is around $10^{16} -10^{17} W/cm^2$ depending on the focusing optical element. (The applied seed oscillator for the amplifier is a standard home-build Ti:sapphire oscillator delivering 20 fs long pulses with 80 MHz repetition rate and with 7 nJ pulse energy.)
The intensity dependence and the presumable saturation of the ionization probability of the free atoms will be measured by detecting the ions and/or electrons. 
The experimental setup can be seen in Fig. 1.
In order to prevent the effects of plasma screening, positive ions are detected throughout the experiments. Single ionized $Rb^+$ ions are collected by the negatively charged cathode of the MCP. It is important that the interaction range must be isolated from all other constant voltages to prevent escaping of the ions. 

Alternatively to ion detection, electrons can be detected as well. In this case a positive charge can be applied to the electrode which is grounded in Fig. 1  In order to check that the atoms are single ionized only and also to measure the density, simultaneous detection of electrons at the position of the grounded electrode is possible as well. In this case the MCP is connected so that even its cathode is positively charged whereas the anode is connected to -2.7 kV higher voltage. Using this arrangement in the case of low densities (in which case the interaction range is within the Debye length of the plasma) equal number of ions and electrons are expected in the case of single ionization. For higher Rb densities the difference of ion-caused current and electron-caused  current gives a possibility to determine the atomic density.
 
The electrons are produced by the 35 fs laser pulses in the high field ionization of the Rb atomic beam of slab like shape from the electrically heated disperser and accelerated by an electrical field of 1000 V on the electrodes A and B into the direction of the cathode (D) of a double chevron michrochannel plate (MCP) trough the drift tube (C). The resulting impulse is observed on the anode (E) of the MCP by an oscilloscope.
A simple rubidium disperser manufactured by ‘Saes getters’ serves as the first version of the source of the atomic beam (see Fig. 2). It contains a suitable compound (chromate) of rubidium which, under appropriate heating by current, emits neutral Rb atoms, while the rest of the components are absorbed by the reducing agent of the dispenser, thus producing a clean source of Rb atoms. Alkali metal dispensers are commonly used as controlled sources of rubidium atoms e.g. for loading magneto-optical traps in quantum optics experiments, e.g. \cite{peter}, and several constructions have been developed to produce directed beams from metal dispensers in this field \cite{roach,comroy}.

  The shape of the dispenser (A) is slab like of about 30 times 1 mm. The metallic slab containing the rubidium compound is heated by electric current. The rubidium atoms released after the dissociation of the compound disperse from the source of temperature of some hundred degrees and distributed in almost  $2\pi$ solid angle. For the aim of the experiments the atoms restricted to a narrow solid angle by two slit apertures positioned to about one centimeter distance.
  
The construction of the rubidium source box can be seen in Figure 2 (except for the side openings which are not indicated here).

 The dispenser and one of the two slit apertures placed in a closed metallic box (B) the top of which contains the second aperture (C). The two slits collimate the emission of the atoms in a narrow region of the vacuum space. The side walls of the two regions of the box have openings in order to avoid the buildup of dense gas phases, atoms emitted by the dispenser wire directly towards the slits can only be emitted. Closely spaced cover plates gather the atoms escaping the box through the side openings. The advantage of using disperser as the source of atomic beam is its flexibility in switching on or off securing its economic long time use.
We have produced well collimated atomic emission profiles with this source with low background densities of rubidium. The atomic density was monitored by measuring the fluorescence excited by resonant laser radiation. The beam of an external cavity diode laser, frequency stabilized close to the strongest resonance line of the 85Rb atoms, was sent through the atomic beam parallel to the slits, and the fluorescence intensity was measured at different positions of the beam. The accumulated profile of the fluorescence can be seen in Figure 3, and a full width at half maximum of about 1.8 mm has been obtained a few mm above the source surface.

 This experimental arrangement placed into a stainless steel vacuum chamber with the ports for entry and exit of the laser pulses, for the optical observations and for the electrical feed troughs. The vessel is pumped by turbo molecular pump to better than $ 10^{-6}$  torr vacuum.

 The field ionization measurements, the next step are now under way.

\section{Theory: Pre-excitation of Rubidium atoms} 

When considering generation of laser plasma with high homogeneity using multi-photon or over-the-barrier ionization (OBI), it is advantageous to find ways for decreasing the threshold conditions for the ionization, especially when generation of a spatially extended plasma is the main goal.  One of the ways to decrease the threshold of  the laser multi-photon ionization (MPI) or OBI is to transfer atomic population from its ground state to one of the excited states before the ionization process, thus significantly decreasing the ionization potential. Of course, such pre-excitation of the atoms must be performed by methods  that are robust and are not too sensitive to the laser pulse parameters including duration, fluctuations in pulse intensity, phase, etc.

One of the most efficient ways to transfer the atomic population to a target state with probability close to unity, is to utilize the method of rapid adiabatic passage (RAP). The word rapid relates to the interaction time that must be shorter than all relaxation times in the system. On the other hand, the variation of the laser parameters must be adiabatically slow to maintain the atom in a given dressed state, \cite{all, berg}.  An approximate condition foradiabaticity is  $\Omega \tau_p >1 $, where $\Omega$  is the peak Rabi frequency, and $\tau_p$  is the duration of the laser pulse. While the relaxation times in gases are in the nano and pico-second  time range, the relaxation processes may be avoided already for laser pulses in the femto- ( tens of femto-) second durations range.  

RAP may be efficiently realized using phase-modulated (frequency-chirped ( FC)) laser pulses, which frequency varies monotonically in time during the laser pulse.  An instructive example of the population transfer produced by FC pulse is the action of such a pulse on a model two-level atom. While action of a laser pulse with constant frequency results in the well-known Rabi oscillations with in average 50 \% population transfer, all the atomic population is transferred from one state to another one when a FC laser pulse is applied, the frequency of which is swept through the resonance with the atomic transition. An important feature of the interaction is its robustness concerning variations of the laser parameters \cite{all, berg}. 

The FC pulses were successfully utilized also in more complicated multilevel atomic or molecular structures \cite{djo1, djo2, djo3, bak1, dem, bak2, djo4, abo}. In Ref. \cite{djo2}, for example,  FC pulses from a diode laser having a central frequency resonant with the D2 line of Rb85  produced a complete population transfer from a ground state to manifold of excited states and back, each time transferring a twice of photon momentum to the atom by a pair of counter-propagating pulses \cite{bak1, dem,  bak2}. The FC pulses were used in \cite{krug} to perform resonance enhanced multi-photon ionization in sodium atoms.

In this communication, we show on the example of Rubidium atom, see Fig.4,  that  pre-excitation of the multilevel atom may be efficiently and robustly performed  by application of FC laser pulses. 
As it is seen from Fig.4, absorption of three photons of radiation of 800 nm wavelength (typical for the Ti:Sa laser)  provides ionization of Rb atom from the ground state 5s. The efficiency of the ionization process may be increased in some extent using intermediate resonances providing resonance enhanced multi-photon ionization, (see e.g. \cite{krug}). However, the threshold of the ionization processes may be drastically decreased if one could transfer the atomic population from the ground level 5s to excited states 5d or 7s, from which only one-photon transition to the vacuum-state provides ionization of the atom.

We have analyzed the possibilities of excitation of the Rb atom by laser pulses having a Gaussian shape with duration equal to 30 fsec, peak intensity in the range of a few $10^{12} W/cm^2$ with a linear frequency chirp having speed in the range of a few  fs$^{-2}$ . Since the duration of the laser pulse is much shorter than the atomic relaxation times, the Schrödinger equation for the probability amplitudes of the working levels of Rb atom in Fig.4, is solved numerically and the dynamics of the atomic populations in the field of the FC pulses is determined, see Fig.5.  
 
As it is clearly seen in Fig.5, a complete population transfer from the ground state 5s to the excited states 5d and 7s is performed by a FC laser pulse. 
The robustness of the excitation process is well seen from the color map in Fig.6: the population of the state 5d after the interaction with the laser pulse remains nearly the same value (close to unity) in a broad range of variation of the peak Rabi frequency and the speed of the positive chirp. 
 
In conclusion, we have shown that it is possible to efficiently transfer all the population of Rb atom from its ground state to the excited states using frequency chirped laser pulses. The presented results of the numerical simulations demonstrate high robustness of the excitation process against variations of the laser pulse parameters that may play a crucial role in generation of a highly homogeneous  spatially extended laser plasma.

In the same AWAKE collaboration one of us  (S. Varr\'o) investigated a different question, namely  the Dirac equation 
of a charged particle interacting with an electromagnetic wave in a medium and found a new 
exact solution \cite{varro}. This problem is significant for the final realization of the AWAKE experiment. 




\begin{figure} 
\scalebox{0.65}{\rotatebox{0}{\includegraphics{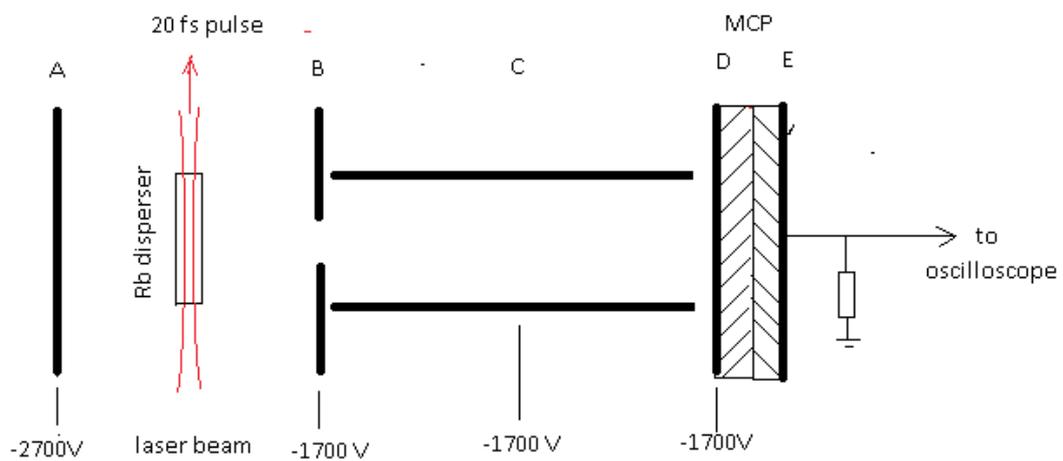}}} 
\vspace*{-2.8cm}
\caption{The experimental setup.}	
\label{egyes}       
\end{figure}

\begin{figure} 
\scalebox{0.5}{\rotatebox{0}{\includegraphics{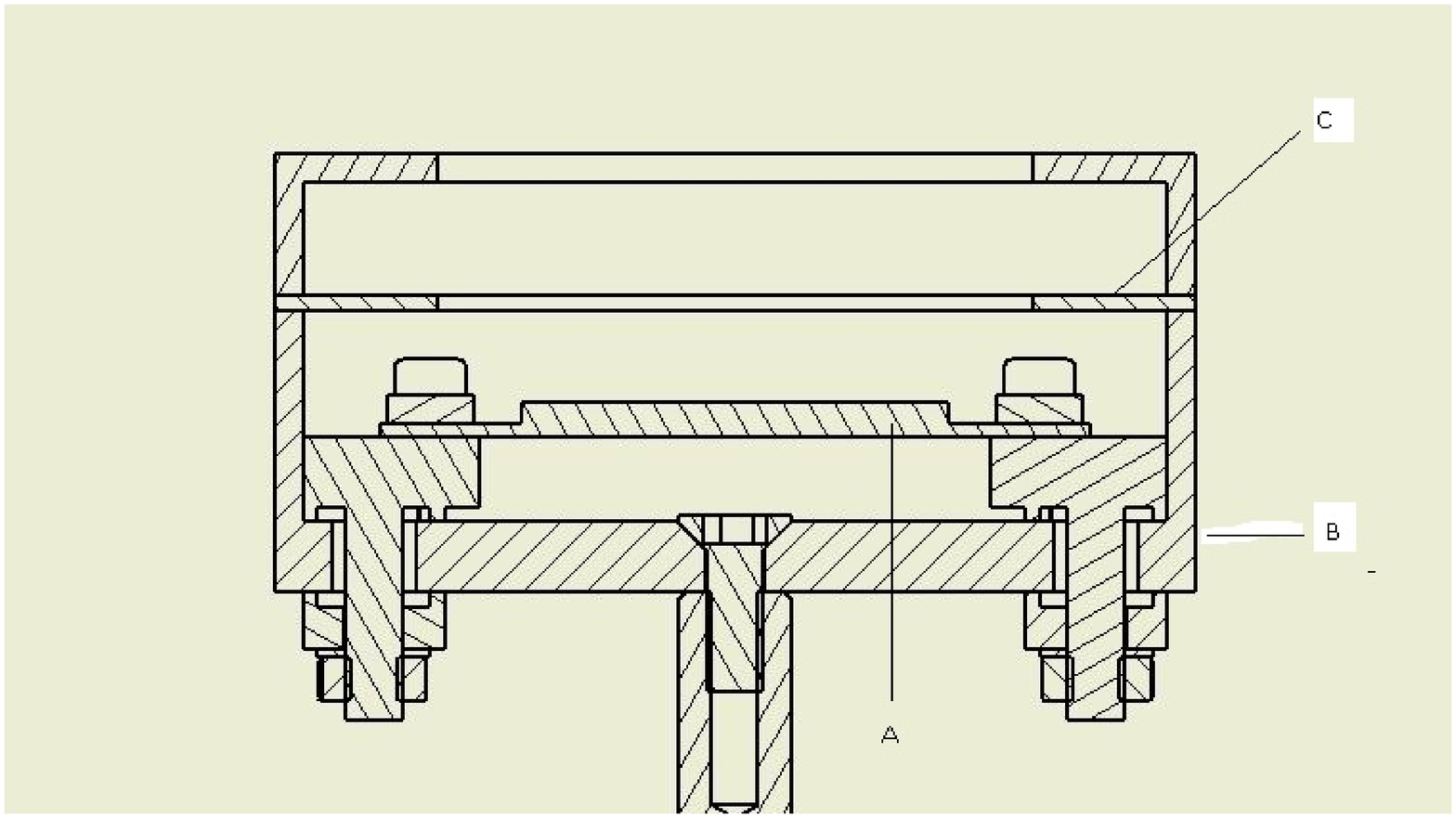}}} 
\caption{The geomtery of the rubidium source box.}	
\label{egyes}       
\end{figure}

\begin{figure} 
\scalebox{0.75}{\rotatebox{0}{\includegraphics{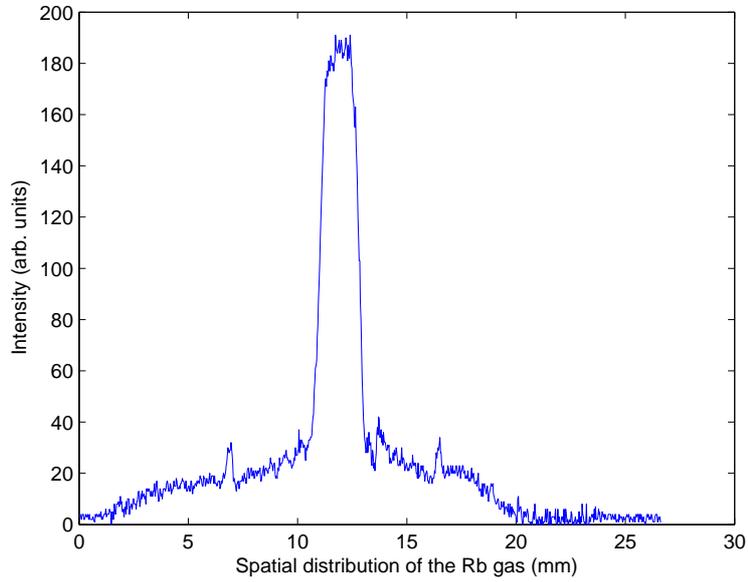}}} 
\caption{The shape of the Rb atomic beam.}	
\label{egyes}       
\end{figure}

\begin{figure} 
\scalebox{0.4}{\rotatebox{0}{\includegraphics{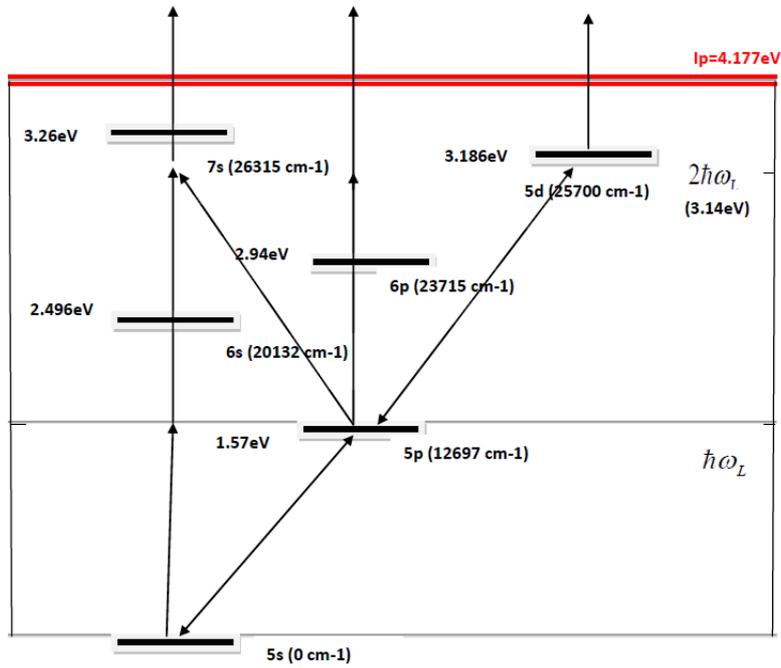}}} 
\caption{The relevant energy levels of Rb atom along with the excitation and ionization scheme.}	
\label{egyes}       
\end{figure}

\begin{figure} 
\scalebox{0.4}{\rotatebox{0}{\includegraphics{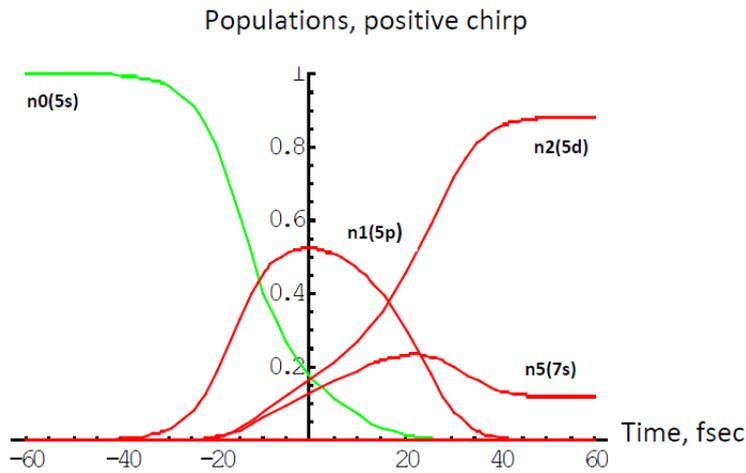}}} 
\caption{Dynamics of the populations of the energy levels of Rb atom in the field of a Gaussian FC laser pulse of 30 fsec duration and positive frequency chirp: the carrier frequency is increasing from the leading to the rear front of the pulse. The pulse  peak intensity is equal to $3x10^{12} W/cm^2$, the speed of the linear chirp is 0.3 fs$^{-2}$.}	
\label{egyes}       
\end{figure}

\begin{figure} 
\scalebox{0.4}{\rotatebox{0}{\includegraphics{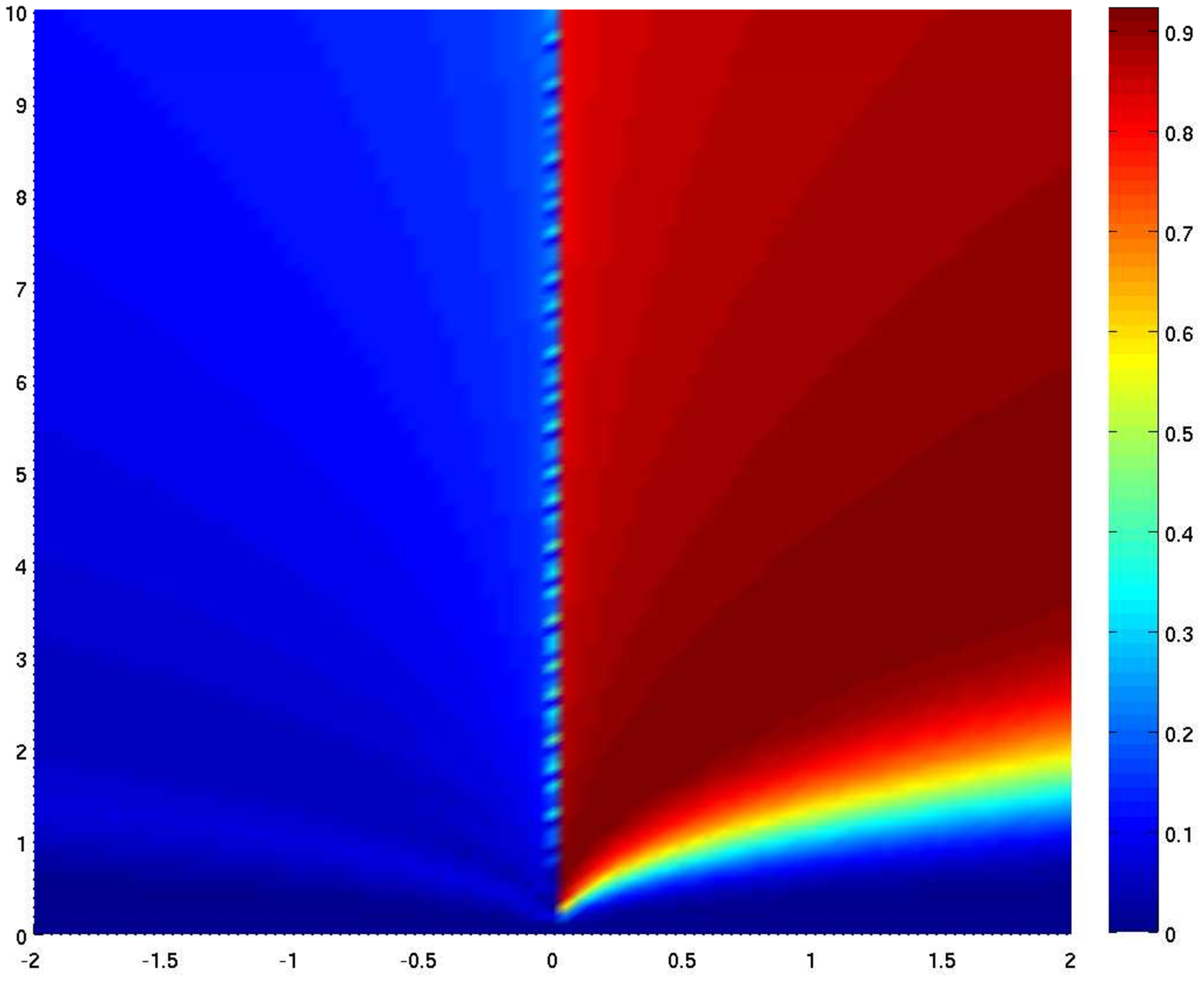}}} 
\caption{Dependence of the final population of the state 5d of the Rb atom on peak Rabi frequency $\Omega_0  $ of the laser pulse (in units of fsec$^{-1}$ ) and 
chirp speed $ \beta $  (in the units of fsec$^{-2}$). }	
\label{egyes}       
\end{figure}


\begin{thebibliography}{00}


 \bibitem{tos} T. Tajima and J. M. Dawson, Phys. Rev. Lett. 43 (1979) 267. 

\bibitem{leh} E. Esarey, C.B. Schroeder, and W.P. Leemans, Rev. of Mod. Phys. 81 (2009) 1229. 

\bibitem{SLAC}  I. Blumenfeld et al., Nature 445 (2007)  741. 

\bibitem{awake} AWAKE Proposal, CERN-SPSC-2013-013 / SPSC-TDR-003. 

\bibitem{prop} A. Pukhov, et al., Principles of self-modulated proton driven plasma wake field acceleration, AIP Conf. Proc. 1507, 103 (2012); doi: 
10.1063/1.4773682. 

\bibitem{peter}  P. D. McDowall, T. Gr\"unzweig, A. Hilliard and M. F. Andersen, Review of Sci. Instr. 83 (2012) 055102.

\bibitem{roach} T. M. Roach and D. Henclewood, J. Vac. Sci. Technol. A 22(6)  (2004) 2384.

\bibitem{comroy} R.S. Conroy, Y. Xiao, M. Vengalattore, W. Rooijakkers and M. Prentiss, Opt. Commun. 226 (2003) 259.

\bibitem{all} L. Allen and J. H. Eberly, Optical Resonance and Two-Level Atoms, (Dover, New York, 1987).

\bibitem{berg} K. Bergmann, H.Theuer, and B. W. Shore, Rev. Mod. Phys. 70 (1998) 1003.

\bibitem{djo1} G.P. Djotyan, J.S Bakos, and  Zs. S\"orlei, Phys. Rev. A, 64 (2001) 013408.

\bibitem{djo2} G.P. Djotyan, J.S. Bakos, G.Demeter, P.N. Ign\'acz, M.A. Kedves, Zs. S\"orlei, J. Szigeti and Z.L. T\'oth, Phys.Rev. A, 68 (2003)  053409.

\bibitem{djo3} G.P. Djotyan, J.S. Bakos and Zs. S\"orlei, Phys.Rev.A, 70 (2004) 063406.

\bibitem{bak1} J.S. Bakos, G.P. Djotyan, P.N. Ignácz, M. A. Kedves, M. Ser\'enyi, Zs. S\"orlei, J. Szigeti and Z. L.T\'oth, Eur. Phys. J. D, 39 (2006) 59.

\bibitem{dem} G. Demeter, G. P. Djotyan, Zs. S\"orlei, and J. S. Bakos, Phys. Rev. A 74 (2006) 013401.

\bibitem{bak2} J.S. Bakos, G.P. Djotyan, P.N. Ign\'acz, M.A. Kedves, M. Ser\'enyi, Zs. S\"orlei, J. Szigeti, and Z.L. T\'oth,  Eur. Phys. J. D 44 (2007) 141.

\bibitem{djo4} G.P. Djotyan, N. Sandor, J.S. Bakos and Zs. S\"orlei, Opt. Express, 19 (2011) 17493.

\bibitem{abo} G.A. Abovyan, G.P. Djotyan and G.Yu. Kryuchkyan, Phys. Rev. A, 85 (2012) 013846.

\bibitem{krug} M. Krug, T. Bayer, M. Wollenhaupt et. al., New J. of Phys.,11 (2009) 105051.

\bibitem{varro} S. Varr\'o, Laser Phys. Lett. 10 (2013) 095301, or arxiv:1305.4370 [quant-ph].  

 \end{thebibliography}
\end{document}